\documentclass{article}
\usepackage{dcase2023,amsmath,url,times,booktabs,tabularx,amsmath,amssymb}
\usepackage{array,multirow}
\usepackage{arydshln}
\usepackage{multicol}
\usepackage{footnote}
\usepackage{float}
\usepackage{cite}
\usepackage{enumitem}
\usepackage{color,soul}

\usepackage[pdftex]{graphicx}

\newcolumntype{C}[1]{>{\centering\arraybackslash}p{#1}}
\newcolumntype{L}[1]{>{\raggedright\arraybackslash}p{#1}}
\newcolumntype{R}[1]{>{\raggedleft\arraybackslash}p{#1}}

\newcommand{\ReduceSpaceUnderFigure}{\vspace{-5mm}}
\newcommand{\ReduceSpaceUnderTable}{\vspace{-5mm}}

\makeatother
\newcommand{\vect}[1]{{\mbox{\boldmath $#1$}}}

\title{DESCRIPTION AND DISCUSSION ON DCASE 2023 CHALLENGE TASK 2: FIRST-SHOT UNSUPERVISED ANOMALOUS SOUND DETECTION FOR MACHINE CONDITION MONITORING}

\name{
Kota Dohi$^{1}$,
Keisuke Imoto$^{2}$,
Noboru Harada$^{3}$,
Daisuke Niizumi$^{3}$,
Yuma Koizumi$^{4}$,
Tomoya Nishida$^{1}$,
}
\secondlinename{
Harsh Purohit$^{1}$,
Ryo Tanabe$^{1}$,
Takashi Endo$^{1}$,
and Yohei Kawaguchi$^{1}$
}
\address{
$^1$ Hitachi, Ltd., Japan, \url{kota.dohi.gr@hitachi.com}\\
$^2$ Doshisha University, Japan, \url{keisuke.imoto@ieee.org}\\
$^3$ NTT Corporation, Japan, \url{noboru.harada.pv@hco.ntt.co.jp}\\
$^4$ Google, Japan, \url{koizumiyuma@google.com}\\
}

\begin{document}

\ninept
\maketitle

\begin{sloppy}

  \begin{abstract}
    We present the task description of the Detection and Classification of Acoustic Scenes and Events (DCASE) 2023 Challenge Task 2: ``First-shot unsupervised anomalous sound detection (ASD) for machine condition monitoring''.
    The main goal is to enable rapid deployment of ASD systems for new kinds of machines without the need for hyperparameter tuning.
    In the past ASD tasks, developed methods tuned hyperparameters for each machine type, as the development and evaluation datasets had the same machine types.
    However, collecting normal and anomalous data as the development dataset can be infeasible in practice.
    In 2023 Task 2, we focus on solving the first-shot problem, which is the challenge of training a model on a completely novel machine type. Specifically, (i) each machine type has only one section (a subset of machine type) and (ii) machine types in the development and evaluation datasets are completely different.
    Analysis of 86 submissions
    from 23 teams revealed that the keys to outperform baselines were:
    1) sampling techniques for dealing with class imbalances across different domains and attributes, 2) generation of synthetic samples for robust detection, and 3) use of multiple large pre-trained models to extract meaningful embeddings for the anomaly detector.
  \end{abstract}


  \section{Introduction}
  \label{sec:intro}

  Anomalous sound detection (ASD)~\cite{koizumi2017neyman, kawaguchi2017how, koizumi2019neyman, kawaguchi2019anomaly, koizumi2019batch, suefusa2020anomalous, purohit2020deep} is the task of identifying whether the sound emitted from a target machine is normal or anomalous. Automatic detection of mechanical failure is essential for the artificial intelligence (AI)--based factory automation. Use of machine sounds for promptly detecting machine anomalies is useful for monitoring a machine's condition.

  One fundamental challenge regarding the application of ASD systems is that anomalous samples for training can be insufficient both in number and type. In 2020, we organized the first ASD task in Detection and Classification of Acoustic Scenes and Event (DCASE) Challenge 2020 Task 2~\cite{Koizumi2020dcase}; ``\textit{unsupervised ASD}'' that was intended to detect unknown anomalous sounds using only normal sound samples as the training data~\cite{koizumi2017neyman, kawaguchi2017how, koizumi2019neyman, kawaguchi2019anomaly, koizumi2019batch, suefusa2020anomalous, purohit2020deep}.

  For the wide-spread application of ASD systems, advanced tasks such as handling of domain shifts should be tackled \cite{Kawaguchi2021}.
  Domain shifts are differences between the source and target domain data caused by a machine's operational conditions or environmental noise.
  Since methods developed in the task in 2020 fail to distinguish normal sounds subject to domain shifts and anomalous sounds, the detection performance of these methods can degrade under domain-shifted conditions.
  To reflect domain-shifted conditions, we organized DCASE 2021 Task 2 ~\cite{Kawaguchi2021}, ``\textit{unsupervised ASD under domain shifted conditions}'' and DCASE Challenge 2022 Task 2~\cite{Dohi2022}, ``\textit{unsupervised ASD applying domain generalization techniques}''. The task in 2021 focused on handling domain shifts using domain adaptation techniques, and the task in 2022 focused on handling domain shifts using domain generalization techniques.

  Previous tasks from 2020 to 2022 had premises such as multiple machine IDs or section IDs for each machine type and the same set of machine types for the development and evaluation datasets.
  As a result, developed methods made use of multiple IDs within a machine type or tuned hyperparameters using normal and anomalous data from the development dataset.
  However, these premises could pose a barrier when attempting to apply methods developed in the past tasks to real-world scenarios, as preparing multiple IDs for each machine type or collecting normal and anomalous data for the development dataset can be time-consuming or even infeasible.


  To solve the problem described above, we designed DCASE Challenge 2023 Task 2, ``\textit{First-Shot Unsupervised Anomalous Sound Detection for Machine Condition Monitoring}''. This task is aimed at developing methods for solving the first-shot problem and rapidly deploying ASD systems, while the task also focuses on developing domain generalization techniques for handling domain shifts. Specifically, only one section is provided for each machine type, and the sets of machine types are completely different between the development and evaluation datasets.

  We received 86 submissions
  from 23 teams. By analyzing these submissions, we found techniques several top-rankers used in common:
  1) sampling techniques for dealing with class imbalances , 2) generation of synthetic samples for robust detection, and 3) use of multiple large pre-trained models to extract meaningful embeddings for the anomaly detector.

  \section{First-shot Unsupervised Anomalous Sound Detection under Domain Shifted Conditions}
  \label{sec:uasd}

  Let the $L$-dimensional time-domain observation $\vect{x}_{i} \in \mathbb{R}^L$ be an audio clip that includes a sound emitted from a machine with a specific ID $i$. The ID serves as a unique identifier that indicates the machine's class on the basis of its model number or other identifying specifications. The goal of the ASD task is to classify the machine as normal or anomalous by computing the anomaly score $\mathcal{A}_{\theta}(\vect{x}_{i})$ by using an anomaly score calculator $\mathcal{A}$ with parameters $\theta$. $\mathcal{A}$ is trained to assign higher scores to anomalous samples and lower scores to normal samples. The input to $\mathcal{A}$ can be the audio clip $\vect{x}_{i}$ or $\vect{x}_{i}$ with additional information such as the ID.
  The machine is classified as anomalous if  $\mathcal{A}_{\theta}(\vect{x}_{i})$ exceeds a pre-defined threshold $\phi$
  \begin{equation}
    \mbox{Decision} = \left\{
    \begin{array}{ll}
      \mbox{Anomaly} & (\mathcal{A}_{\theta}(\vect{x}_{i}) > \phi) \\
      \mbox{Normal}  & (\mbox{otherwise}).
    \end{array}
    \right.
    \label{eq:det}
  \end{equation}
  The primary difficulty in this task is to train $\mathcal{A}$ using only normal sounds (unsupervised ASD). The DCASE 2020 Challenge Task 2 was designed to address this issue.

  In real-world scenarios, the domain-shift problem also needs to be solved. Domain shifts are variations in conditions between training and testing phases that impact the distribution of normal sound data.
  These shifts can arise from differences in operating speed, machine load, viscosity, heating temperature, environmental noise, signal-to-noise ratio, and other factors.
  Two domains, \textbf{source domain} and \textbf{target domain}, are defined:
  the former refers to the original condition with sufficient training data and the latter refers to another condition with only a few samples.
  The 2021 Task 2 aimed to develop domain adaptation techniques, assuming the domain information (source/target) of each sample is known. However, in practice, obtaining domain information is challenging due to the difficulty in detecting domain shifts.

  To address the challenges of applying domain adaptation techniques in real-world scenarios, the 2022 Task 2 focused on developing domain generalization techniques. Domain generalization techniques for ASD aim at detecting anomalies from different domains with a single threshold. These techniques, unlike domain adaptation techniques, do not require detection of domain shifts or adaptation of the model during the testing phase.

  Although several novel ASD methods have been proposed in past tasks, we have recognized that their application in real-world scenarios remains challenging. This is because certain assumptions in previous tasks may not hold in practice. One such assumption is that participants were allowed to tune the hyperparameters of the model by using the test data of the development dataset. However, this is often infeasible in real-world applications where the machine type can be completely new or the amount of test data can be insufficient for tuning hyperparameters.
  Another assumption is the existence of multiple IDs for a machine type. This assumption has facilitated the development of outlier exposure approaches \cite{giri2020self}, where sound clips from different machines are used as anomalies. However, in many practical cases, the number of machines for a machine type can be limited. This limitation arises because the customers may not possess multiple machines of the same machine type, or they may initially plan to install the system for only a few machines. As a result, the developed methods in the previous tasks may not be immediately applicable in practice.

  To overcome these new challenges, the organizers designed the 2023 Task 2 with two main features:
  (i) completely different set of machine types between the development and evaluation dataset and (ii) Only one section for each machine type.
  Because the machine types are completely different between the development and evaluation dataset, tuning hyperparameters using the test data from the development dataset is no longer feasible. Furthermore, since only one section is available for each machine type, multiple IDs within a machine type cannot be used.
  As a result, participants are expected to develop ASD methods without tuning hyperparameters using the test data and without relying on multiple IDs within a machine type.
  We name these challenges the ``first-shot problem'', as these challenges replicate practical cases where the ASD system has to be deployed for a novel machine type or with a limited number of example measurements.

  \section{Task Setup}
  \label{sec:task}

  \subsection{Dataset}
  \label{sec:dataset}

  The data for this task comprises three datasets: \textbf{development dataset}, \textbf{additional training dataset}, and \textbf{evaluation dataset}.
  Each dataset includes seven machine types, with one section per machine type. \textbf{Machine type} means the type of machine such as fan, gearbox, bearing, etc.  \textbf{Section} is a subset or whole data within each machine type.

  Each recording is a single-channel audio with a duration of 6 to 18 s and a sampling rate of 16 kHz. We mixed machine sounds recorded at laboratories and environmental noise samples recorded at factories and in the suburbs to create each sample in the dataset.
  For the details of the recording procedure, please refer to the papers on ToyADMOS2~\cite{harada2021toyadmos2} and MIMII DG~\cite{Dohi2022}.

  The \textbf{development dataset} consists of seven machine types (fan, gearbox, bearing, slide rail, ToyCar, ToyTrain), and each machine type has one section that contains a complete set of the training and test data.
  Each section provides
  (i) 990 normal clips from a source domain for training,
  (ii) 10 normal clips from a target domain for training, and
  (iii) 100 normal clips and 100 anomalous clips from both domains for the test. We provided domain information (source/target) in the test data for the convenience of participants. Attributes that represent operational or environmental conditions are also provided in the file names and attribute csvs.

  The \textbf{additional training dataset} provides novel seven machine types (Vacuum, ToyTank, ToyNscale, ToyDrone, bandsaw, grinder, shaker).
  Each section consists of
  (i) 990 normal clips in a source domain for training and
  (ii) 10 normal clips in a target domain for training.
  Attributes are provided in this dataset.

  The \textbf{evaluation dataset} provides the same machine types as the additional training dataset.
  Each section consists of 200 test clips, none of which have a condition label (i.e., normal or anomaly) or the domain information. Attributes are not provided.

  The data for this task differs from the 2022 version in two main aspects: reduced number of sections per machine type (from six in 2022 to one in this task) and a completely different set of machine types between the development and evaluation datasets.
  As a result, participants are required to train a model for a novel machine type using only one section for each machine type and without hyperparameter tuning using the development dataset.

  \subsection{Evaluation metrics}
  \label{sec:metrics}

  For evaluation, the area under the receiver operating characteristic curve (AUC) was employed as a metric to assess the overall detection performance, while the partial AUC (pAUC) was utilized to measure performance in a low false-positive rate (FPR) range $\left[ 0, p \right]$.
  In this task, we used $p=0.1$.
  In domain generalization task, the AUC for each domain and pAUC for each section are calculated as
  \begin{equation}
    {\rm AUC}_{m, n, d} = \frac{1}{N^{-}_{d}N^{+}_{n}} \sum_{i=1}^{N^{-}_{d}} \sum_{j=1}^{N^{+}_{n}}
    \mathcal{H} (\mathcal{A}_{\theta} (x_{j}^{+}) - \mathcal{A}_{\theta} (x_{i}^{-})),
  \end{equation}
  \begin{equation}
    {\rm pAUC}_{m, n} = \frac{1}{\lfloor p N^{-}_{n} \rfloor N^{+}_{n}} \sum_{i=1}^{\lfloor p N^{-}_{n} \rfloor N^{+}_{n}} \sum_{j=1}^{N^{+}_{n}}
    \mathcal{H} (\mathcal{A}_{\theta} (x_{j}^{+}) - \mathcal{A}_{\theta} (x_{i}^{-})),
  \end{equation}
  where $m$ represents the index of a machine type,
  $n$ represents the index of a section,
  $d = \{ {\rm source}, {\rm target} \}$ represents a domain,
  $\lfloor \cdot \rfloor$ is the flooring function,
  and $\mathcal{H} (x)$ returns 1 when $x > 0$ and 0 otherwise.
  Here, $\{x^{-}_{i}\}_{i=1}^{N^{-}_{d}}$ are normal test clips in domain $d$ in section $n$
  and $\{x_{j}^{+}\}_{j=1}^{N^{+}_{n}}$ are anomalous test clips in section $n$ in machine type $m$.
  The $N^{-}_{d}$ is the number of normal test clips in domain $d$,
  $N^{-}_{n}$ is the number of normal test clips in section $n$,
  and $N^{+}_{n}$ is the number of anomalous test clips in section $n$.

  The official score $\Omega$ is given by the harmonic mean of the AUC and pAUC scores over all machine types and sections:
  \begin{eqnarray}
    \Omega &=& h \left\{ {\rm AUC}_{m, n, d}, \ {\rm pAUC}_{m, n} \quad | \quad \right. \nonumber \\
    && \left. m \in \mathcal{M}, \  n \in \mathcal{S}(m), \ d \in \{ {\rm source}, {\rm target} \} \right\},
  \end{eqnarray}
  where $h\left\{\cdot\right\}$ represents the harmonic mean (over all machine types, sections, and domains),
  $\mathcal{M}$ represents the set of machine types,
  and $\mathcal{S}(m)$ represents the set of sections for machine type $m$.


  \subsection{Baseline systems and results}
  \label{sec:baseline}

  The organizers provided an Autoencoder (AE)-based baseline system with two different ways of calculating the anomaly scores.
  We present the baseline system and its detection performance. For details, please refer to \cite{Harada2023}.

  \subsubsection{Autoencoder-based baseline}
  First, the log-mel-spectrogram of the input $X = \{X_k\}_{k = 1}^K$ is calculated,
  where $X_k \in \mathbb{R}^F$, and $F$ and $K$ are the number of mel-filters and time-frames, respectively.
  Then, the acoustic feature at $k$ is obtained by concatenating consecutive frames of the log-mel-spectrogram as $\psi_k = (X_k, \cdots, X_{k + P - 1}) \in \mathbb{R}^D$,
  where $D = P \times F$, and $P$ is the number of frames of the context window.



  \begin{figure*}[t]
    \begin{center}
      \includegraphics[width=1.0\hsize,clip]{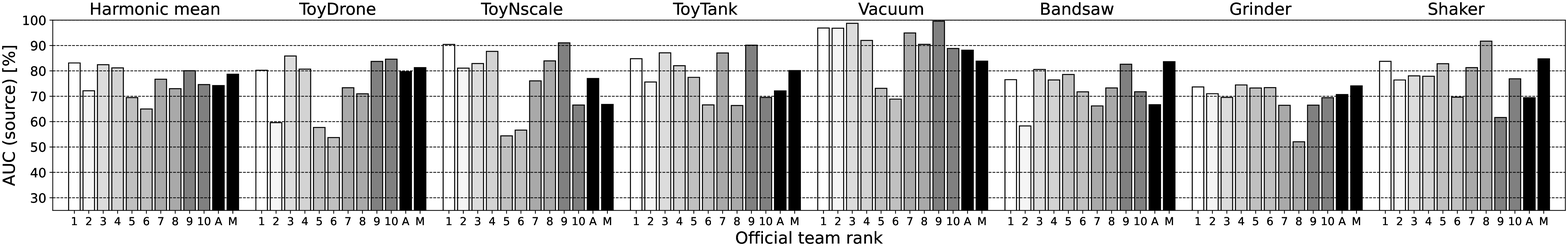}\\
      \includegraphics[width=1.0\hsize,clip]{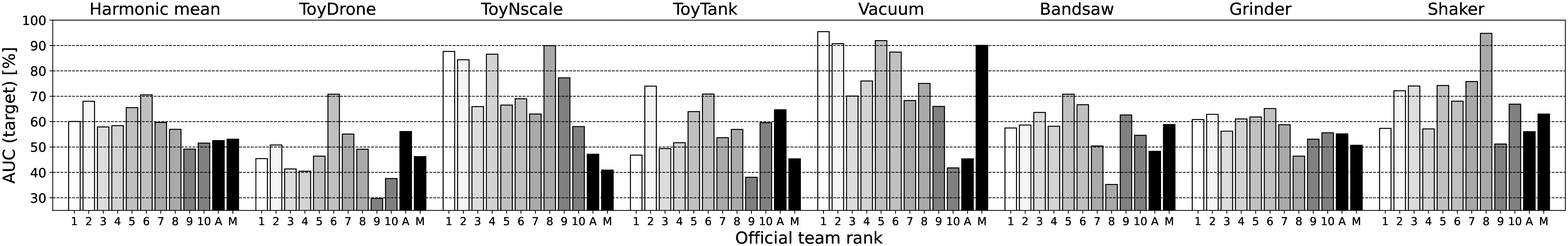}\\
      \ReduceSpaceUnderFigure
      \caption{Evaluation results of top 10 teams in the ranking. Average source-domain AUC (Top) and target-domain AUC
        (bottom) for each machine type. Label ``A'' and ``M'' on the x-axis denote simple Autoencoder mode and selective Mahalanobis mode, respectively.}
      \label{fig:aucs}
    \end{center}
  \end{figure*}

  \begin{figure}[t]
    \begin{center}
      \includegraphics[width=0.9\hsize,clip]{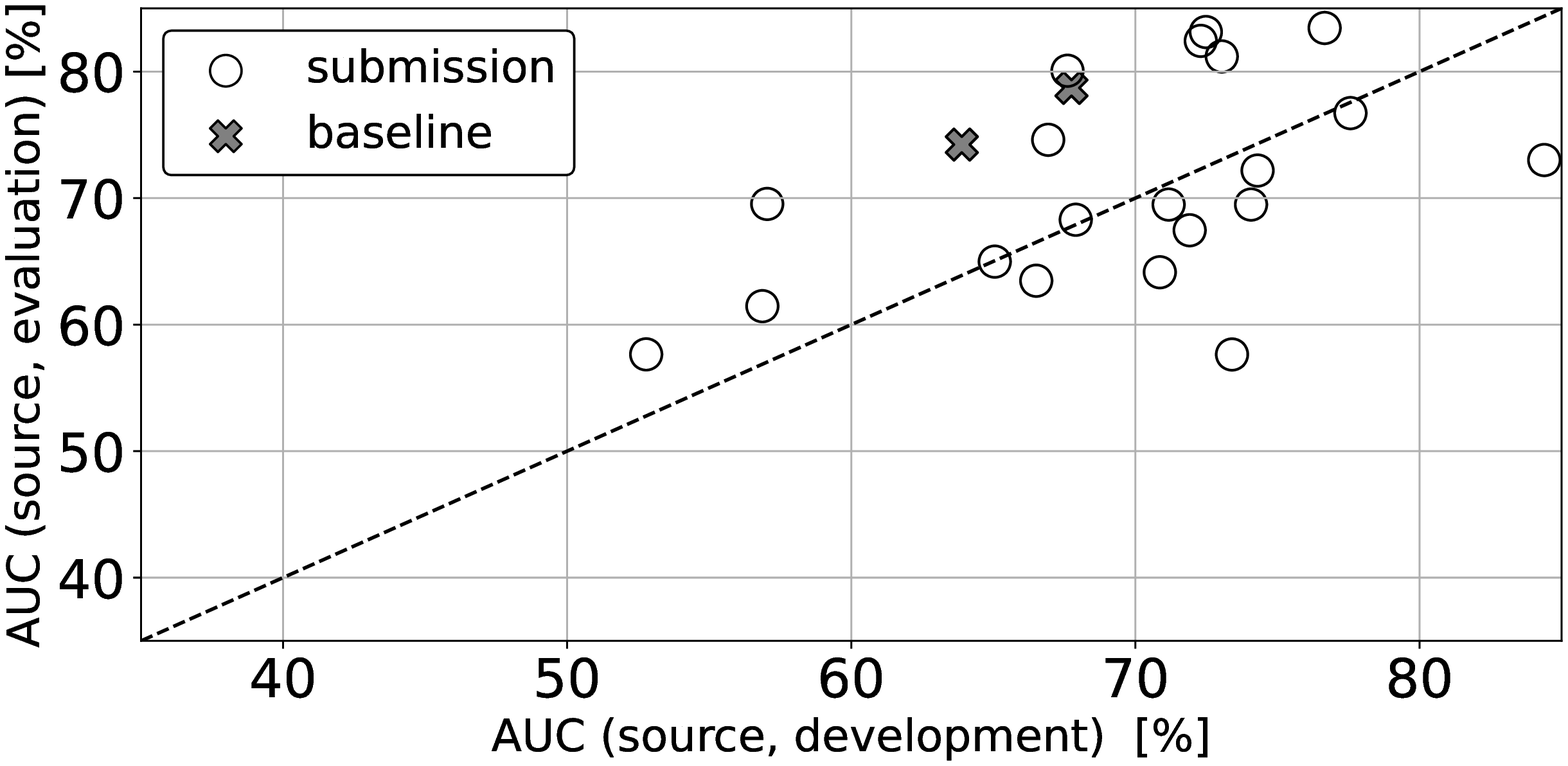}\\
      \ReduceSpaceUnderFigure
      \caption{Comparison of average source-domain AUC for the development dataset and evaluation dataset across teams.}
      \label{fig:source}
    \end{center}
  \end{figure}

  \begin{figure}[t]
    \begin{center}
      \includegraphics[width=0.9\hsize,clip]{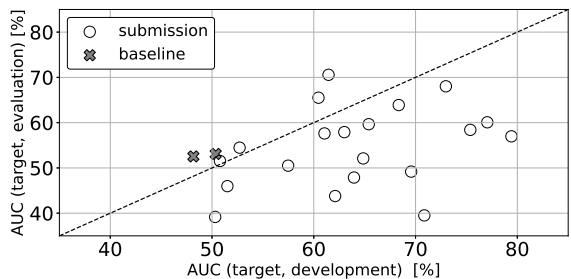}\\
      \ReduceSpaceUnderFigure
      \caption{Comparison of average target-domain AUC for the development dataset and evaluation dataset across teams.}
      \label{fig:target}
    \end{center}
  \end{figure}
  \subsubsection{Simple Autoencoder mode}
  In this mode, the anomaly score is calculated as
  \begin{equation}
    A_{\theta}(X) = \frac{1}{DK} \sum_{k = 1}^K \| \psi_k - r_{\theta}(\psi_k) \|_{2}^{2},
  \end{equation}
  where $r_{\theta}$ is the vector reconstructed by the AE, and $\| \cdot \|_2$ is $\ell_2$ norm.

  \subsubsection{Selective Mahalanobis mode}
  In this mode, the Mahalanobis distance between the observed sound and reconstructed sound is used to calculate the anomaly score. The anomaly score is given as
  \begin{equation}
    A_{\theta}(X) = \frac{1}{DK} \sum_{k = 1}^K min\{ D_s (\psi_k, r_{\theta}(\psi_k)), D_t (\psi_k, r_{\theta}(\psi_k))\},
  \end{equation}
  \begin{equation}
    D_s(\cdot) = Mahalanobis(\psi_k, r_{\theta}(\psi_k), \Sigma_s^{-1}),
  \end{equation}
  \begin{equation}
    D_t(\cdot) = Mahalanobis(\psi_k, r_{\theta}(\psi_k), \Sigma_t^{-1}),
  \end{equation}
  where $\Sigma_s^{-1}$ and $\Sigma_t^{-1}$ are the covariance matrices calculated with the source domain data and target domain data of each section, respectively.

  \subsubsection{Results}
  \label{sec:results}

  \setlength{\tabcolsep}{1mm}
  \begin{table}[]
    \begin{center}
      \caption{Results with Simple Autoencoder mode}
      \label{tab:ae_results}
      \scriptsize
      \begin{tabular}{c c c c p{1pt} c c}
        \hline
        Machine type                 &
        Section                      &
        \multicolumn{2}{c}{AUC [\%]} &    &
        \multicolumn{2}{c}{pAUC [\%]}                                                                 \\
        \cline{3-4} \cline{6-7}
                                     &    &
        \multicolumn{1}{c}{Source}   &
        \multicolumn{1}{c}{Target}   &    &                                                           \\
        \hline
        ToyCar
                                     & 00 & $70.10 \pm 0.46$ & $46.89 \pm 2.67$ &  & $52.47 \pm 1.28$ \\

        ToyTrain
                                     & 00 & $57.93 \pm 2.12$ & $57.02 \pm 0.79$ &  & $48.57 \pm 0.32$ \\

        bearing
                                     & 00 & $65.92 \pm 0.73$ & $55.75 \pm 0.76$ &  & $50.42 \pm 0.79$ \\

        fan
                                     & 00 & $80.19 \pm 2.43$ & $36.18 \pm 3.71$ &  & $59.04 \pm 1.24$ \\

        gearbox
                                     & 00 & $60.31 \pm 0.56$ & $60.69 \pm 0.63$ &  & $53.22 \pm 0.60$ \\

        slider
                                     & 00 & $70.31 \pm 0.20$ & $48.77 \pm 0.12$ &  & $56.37 \pm 0.31$ \\

        valve
                                     & 00 & $55.35 \pm 1.18$ & $50.69 \pm 1.12$ &  & $51.18 \pm 0.35$ \\
        \hline
      \end{tabular}
      \ReduceSpaceUnderTable
    \end{center}
  \end{table}

  \setlength{\tabcolsep}{1mm}
  \begin{table}[]
    \begin{center}
      \caption{Results with Selective Mahalanobis mode}
      \label{tab:mahala_results}
      \scriptsize
      \begin{tabular}{c c c c p{1pt} c c}
        \hline
        Machine type                 &
        Section                      &
        \multicolumn{2}{c}{AUC [\%]} &    &
        \multicolumn{2}{c}{pAUC [\%]}                                                                 \\
        \cline{3-4} \cline{6-7}
                                     &    &
        \multicolumn{1}{c}{Source}   &
        \multicolumn{1}{c}{Target}   &    &                                                           \\
        \hline
        ToyCar
                                     & 00 & $74.53 \pm 1.55$ & $43.42 \pm 2.53$ &  & $49.18 \pm 0.49$ \\

        ToyTrain
                                     & 00 & $55.98 \pm 2.41$ & $42.45 \pm 1.06$ &  & $48.13 \pm 0.17$ \\

        bearing
                                     & 00 & $65.16 \pm 0.76$ & $55.28 \pm 0.57$ &  & $51.37 \pm 0.81$ \\

        fan
                                     & 00 & $87.10 \pm 2.20$ & $45.98 \pm 4.43$ &  & $59.33 \pm 0.90$ \\

        gearbox
                                     & 00 & $71.88 \pm 0.66$ & $70.78 \pm 0.62$ &  & $54.34 \pm 0.30$ \\

        slider
                                     & 00 & $84.02 \pm 1.10$ & $73.29 \pm 0.60$ &  & $54.72 \pm 0.25$ \\

        valve
                                     & 00 & $56.31 \pm 1.38$ & $51.40 \pm 0.40$ &  & $51.08 \pm 0.13$ \\
        \hline
      \end{tabular}
      \ReduceSpaceUnderTable
    \end{center}
  \end{table}

  The AUC and pAUC for each machine type are shown in Tables \ref{tab:ae_results} and \ref{tab:mahala_results}. The results are average of five independent runs.

  \section{Challenge Results}
  \vspace{-0.5mm}
  We received 86 submissions from 23 teams. Eleven teams outperformed the simple Autoencoder baseline, and eight teams outperformed the selective Mahalanobis baseline. The number of teams was significantly fewer than for the task in 2022, where 22 out of 31 teams outperformed the baselines. This observation suggests that the new features in this year's task, such as having only one section for each machine type and novel machine types in the evaluation dataset, have increased the task's difficulty level. Despite these challenges, several top-ranked teams significantly outperformed the baselines. Figure \ref{fig:aucs} illustrates the harmonic means of the AUCs for the top 10 teams.
  Notably, all eight teams that outperformed the baselines in the official scores also surpassed the baselines in the harmonic mean of the AUCs in the target domain. This indicates that higher AUCs in the target domain were crucial for higher ranks.

  Since the task this year focused on developing ASD methods that work for novel machine types, we compared the AUCs between the development and evaluation datasets. Figure \ref{fig:source} shows the AUCs from the top 20 teams for the source domain, while Figure \ref{fig:target} displays the AUCs for the target domain. From Figure \ref{fig:source}, it can be observed that approximately half of the teams achieved higher source-domain AUCs in the evaluation dataset compared to the development dataset. This indicates that, with a sufficient amount of training data, detection for a novel machine type can be possible without significant degradation in performance. However, Figure \ref{fig:target} reveals that the target-domain AUCs were lower in the evaluation dataset for most teams. This underscores the difficulty of dealing with domain shifts for novel machine types. The lower AUCs observed in the evaluation dataset for the target domain can be attributed to the fact that the variations induced by domain shifts can differ significantly for each machine type.
  In this case, when domain generalization techniques are developed for maximizing the AUCs in the development dataset, using the same techniques for the evaluation dataset will degrade the performance.
  Addressing these variations becomes more challenging when only a limited number of samples are available, further complicating the problem.

  We summarize approaches used by top-ranked teams in the following.\\

  \vspace{-3mm}
  \noindent
  \textbf{a. Oversampling for imbalance compensation}

  Because the number of samples in the datasets is imbalanced across domains and attributes, compensating for these class imbalances can improve the detection performance. The 6th team \cite{ZhouSHNU2023} duplicated samples from classes with fewer samples, while the 1st and 2nd teams\cite {JieIESEFPT2023, LvHUAKONG2023} oversampled target-domain data using SMOTE \cite{Chawla2002}. These approaches are only seen among top-rankers, and can be one of the key factors for outperforming the baselines.\\

  \noindent
  \textbf{b. Synthetic data generation for robust detection}

  Synthetic data can be utilized to accurately model the distribution of normal data and enhance the robustness of the detection model. The 1st, 4th, 5th, 10th, and 19th teams employed Mixup \cite{zhang2018mixup} including its variants \cite{JieIESEFPT2023, WilkinghoffFKIE2023, BaiJLESS2023, JiaJunHFUU2023, FujimuraNU2023}, and obtained higher source-domain AUCs. Other papers used other data augmentation techniques such as speed perturbation, noise injection, and pitch shift \cite{LvHUAKONG2023, ZhouSHNU2023, GuanHEU2023}. The treatment of generated synthetic data varies among teams. While the 4th team \cite{WilkinghoffFKIE2023} treated them as anomalous samples that belong to a new class, the 1st and 5th teams \cite{JieIESEFPT2023, BaiJLESS2023} treated them as normal samples. Mixup can be one of the key factors for outperforming the baselines, as this technique was used by several top-rankers and teams that achieved higher source-domain AUCs. \\

  \noindent
  \textbf{c. Attribute ID classification using pre-trained models}

  Although only one section was provided for each machine type, attributes were included in the development and additional training dataset. As a result, many participants trained attribute classifiers or machine type classifiers to obtain embeddings that could be used for outlier detectors \cite{JieIESEFPT2023, LvHUAKONG2023, JiangTHUEE2023, WilkinghoffFKIE2023, ZhouSHNU2023, DuNERCSLIP2023}. For the outlier detector, k-nearest neighbors algorithm (kNN) was used by most of the teams.

  Pre-trained models are used \cite{LvHUAKONG2023, JiangTHUEE2023, LiuCQUPT2023} for attribute classifiers or machine type classifiers. Although pre-trained models have been used by participants in previous tasks, the 2nd and 3rd teams \cite{LvHUAKONG2023, JiangTHUEE2023} are the first teams that used multiple large pre-trained models to achieve higher official scores. These pre-trained models were fine-tuned with classification objectives, i.e., attribute or machine type classification. \\

  \noindent
  \textbf{d. Other novel approaches}

  The 3rd team \cite{JiangTHUEE2023} grouped machine types into several categories so that generalization ability on novel machine types can be obtained. The 7th team \cite{GuanHEU2023} used AudioLDM \cite{liu2023audioldm}, a text-to-audio model, to generate pseudo anomalous sounds from the text input.

  \section{Conclusion}
  This paper presented an overview of the task and analysis of the solutions submitted to DCASE 2023 Challenge Task 2.
  The task was aimed to develop an ASD system that works for a novel machine type with a single section for each machine type. Analysis of the submission revealed that, for novel machine types, detection in the target domain can be of significant difficulty compared to the source domain. The analysis also revealed useful methods for outperforming the baselines: 1) sampling techniques for dealing with class imbalances, 2) generation of synthetic samples by mix-up and its variants, and 3) use of multiple large pre-trained models for attribute ID classification.

  \bibliographystyle{IEEEtran}
  \bibliography{refs}

\end{sloppy}
\end{document}